\documentstyle[12pt]{article}
\def\beq{\begin{equation}}
\def\eeq{\end{equation}}
\def\bea{\begin{eqnarray}}
\def\eea{\end{eqnarray}}

\catcode`\@=11
\def\marginnote#1{}
\def\ifmath#1{\relax\ifmmode #1\else $#1$\fi}

\def\U{{\cal U}}

\def\bold#1{\setbox0=\hbox{$#1$}%
     \kern-.025em\copy0\kern-\wd0
     \kern.05em\copy0\kern-\wd0
     \kern-.025em\raise.0433em\box0 }

\def\GENITEM#1;#2{\par\vskip6pt \hangafter=0 \hangindent=#1
   \Textindent{$ #2$ }\ignorespaces}

\newcount\hour
\newcount\minute
\newtoks\amorpm
\hour=\time\divide\hour by60
\minute=\time{\multiply\hour by60 \global\advance\minute by-
\hour}
\edef\standardtime{{\ifnum\hour<12 \global\amorpm={am}%
    \else\global\amorpm={pm}\advance\hour by-12 \fi
    \ifnum\hour=0 \hour=12 \fi
    \number\hour:\ifnum\minute<100\fi\number\minute\the\amorpm}}
\edef\militarytime{\number\hour:\ifnum\minute<100\fi\number\minute}
\def\draftlabel#1{{\@bsphack\if@filesw {\let\thepage\relax
  \xdef\@gtempa{\write\@auxout{\string
    \newlabel{#1}{{\@currentlabel}{\thepage}}}}}\@gtempa
    \if@nobreak \ifvmode\nobreak\fi\fi\fi\@esphack}
     \gdef\@eqnlabel{#1}}
\def\@eqnlabel{}
\def\@vacuum{}
\def\draftmarginnote#1{\marginpar{\raggedright\scriptsize\tt#1}}
\def\draft{\oddsidemargin -.5truein
        \def\@oddfoot{\sl preliminary draft \hfil
        \rm\thepage\hfil\sl\today\quad\militarytime}
        \let\@evenfoot\@oddfoot \overfullrule 3pt
        \let\label=\draftlabel
        \let\marginnote=\draftmarginnote

\def\@eqnnum{(\theequation)\rlap{\kern\marginparsep\tt\@eqnlabel}%
\global\let\@eqnlabel\@vacuum}  }
\def\preprint{\twocolumn\sloppy\flushbottom\parindent 1em
        \leftmargini 2em\leftmarginv .5em\leftmarginvi .5em
        \oddsidemargin -.5in    \evensidemargin -.5in
        \columnsep 15mm \footheight 0pt
        \textwidth 250mmin      \topmargin  -.4in
        \headheight 12pt \topskip .4in
        \textheight 175mm
        \footskip 0pt

\def\@oddhead{\thepage\hfil\addtocounter{page}{1}\thepage}
        \let\@evenhead\@oddhead \def\@oddfoot{} \def\@evenfoot{}
}
\def\titlepage{\@restonecolfalse\if@twocolumn\@restonecoltrue\o
necolumn
     \else \newpage \fi \thispagestyle{empty}\c@page\z@
        \def\thefootnote{\fnsymbol{footnote}} }
\def\endtitlepage{\if@restonecol\twocolumn \else  \fi
        \def\thefootnote{\arabic{footnote}}
        \setcounter{footnote}{0}}  
\catcode`@=12
\relax
\def\be{\begin{equation}}
\def\ee{\end{equation}}
\def\bea{\begin{eqnarray}}
\def\eea{\end{eqnarray}}

\def\mst11{m_{\;\widetilde{t}_{1}}}

\def\mst22{m_{\;\widetilde{t}_{2}}}
\def\mst12{m_{\;\widetilde{t}_{1,2}}}

\def\msb11{m_{\;\widetilde{b}_{1}}}
\def\msb22{m_{\;\widetilde{b}_{2}}}
\def\msb12{m_{\;\widetilde{b}_{1,2}}}

\def\mwidetilde2{\widetilde{m}^{2}}

\relax

\begin{document}
\input epsf

\topmargin-2.5cm
%
\begin{titlepage}
\begin{flushright}
hep-ph/0112261 \\
DFPD-A/01/57
\end{flushright}
\vskip 0.1in
\begin{center}
{\Large\bf Non-Gaussianity from Inflation}

\vskip .5in
{\large\bf N. Bartolo}$^{1,2,3}$,
{\large \bf S. Matarrese}$^{1,2}$ 
{\large and}
{\large\bf A. Riotto}$^{2}$

\vskip0.7cm
$^{1}${\it Dipartimento di Fisica di Padova ``G. Galilei''}

\vskip 0.1cm

{\it Via Marzolo 8, Padova I-35131, Italy}

\vskip 0.2cm
       
${^2}${\it INFN, Sezione di Padova}

\vskip 0.1cm

{\it Via Marzolo 8, Padova I-35131, Italy}

\vskip 0.2cm

${^3}${\it Astronomy Centre, University of Sussex}

\vskip 0.1cm

{\it Falmer, Brighton, BN1 9QJ, U.K.}

\end{center}

\vskip 1cm

\begin{center}
{\bf Abstract} 
\end{center}
\vskip 0.5cm
Correlated adiabatic and isocurvature perturbation modes are  
produced during inflation through an oscillation mechanism 
when extra scalar degrees of freedom
-other than the inflaton field are present. We show that 
this correlation generically 
leads to sizeable non-Gaussian features 
both in the adiabatic 
and isocurvature perturbations. The non-Gaussianity is first generated 
by large non-linearities in some scalar sector 
and then efficiently transferred to the inflaton sector by the oscillation 
process. We compute the cosmic microwave background angular bispectrum, 
providing a characteristic feature of such inflationary non-Gaussianity,
which might be detected by upcoming satellite experiments.
\begin{quote}

\end{quote}
\vskip1.cm
\begin{flushleft}
March 2002 \\
\end{flushleft}

\end{titlepage}
\setcounter{footnote}{0}
\setcounter{page}{0}

\newpage
%
\baselineskip=24pt
\noindent

\section{Introduction}

It is generally believed that inflation provides the causal 
mechanism to seed structure formation in the Universe. 
One of the most interesting aspects of these primordial perturbations is 
their statistical nature. The simplest and most generally accepted idea is 
that these primordial perturbations were Gaussian distributed. 
However, this issue is far from being settled: there is still 
ample room for some level of non-Gaussianity in the initial conditions. 

One way of parametrizing the possible presence of non-Gaussianity in the 
primordial gravitational potential $\Phi$ is to expand it
in the following way
\cite{NL, FRS, Vetal, ks}
\begin{equation} 
\Phi = \varphi + f_{\rm NL} (\varphi^2 - \langle \varphi^2 \rangle) + 
{\cal O}\left(f_{\rm NL}^2 \right) \;, 
\end{equation}
where $\varphi$ is a zero-mean Gaussian random field and $f_{\rm NL}$
is an expansion parameter which can be observationally constrained. 

It is commonly believed that primordial perturbations generated during 
inflation are necessarily adiabatic and Gaussian. Although this is 
essentially the case for the simplest model, where a single inflaton field 
undergoes a slow-roll transition, the range of possibilities is actually 
much wider and more interesting than such a standard lore may tell. 
Even in the case of a single, slowly rolling inflaton field, it has been shown 
that the effect of field non-linearities and their backreaction on the 
underlying geometry is to generate a small, but calculable, non-Gaussianity 
\cite{FRS, Getal, wk, GM}. 
The non-Gaussianity, or non-linearity, parameter $f_{\rm NL}$ can be expressed 
in terms of the standard slow-roll parameters $\epsilon$ and $\eta$ as 
$f_{\rm NL} \sim 3\epsilon -2 \eta$ \cite{Getal, wk, GM}. 
Since the slow-roll parameters have to be much smaller than unity for 
inflation to occur, the typical value of $f_{\rm NL}$ in single-field 
inflationary models is inevitably tiny. These constraints can be partially 
relaxed  if the inflaton potential contains `features' in that part 
corresponding to the last $\sim 60$ e-foldings \cite{features, KBHP, wk}. 

The common belief that non-Gaussianity of inflation generated perturbations 
is small comes from this theoretical argument applied to
single-field models of inflation. 
On the other hand the presence of non-Gaussianity is only  
mildly constrained by observations. Let us  focus on the evidence for Gaussian 
primordial fluctuations coming from the analysis of primary anisotropies 
of the Cosmic Microwave Background (CMB), as these certainly provide the most 
direct probe of initial conditions and the most efficient way 
to look for non-Gaussianity of the type described by Eq. (1) \cite{Vetal}.   
Recent analyses of the angular bispectrum from 4-year COBE data \cite{ketal} 
yield a weak upper limit, $\vert f_{\rm NL}\vert < 1.5 \times 10^3$.
The analysis of the diagonal angular 
bispectrum of the Maxima dataset \cite{setal} also provides a very 
weak constraint: $\vert f_{\rm NL} \vert < 2330$. 
According to Komatsu and Spergel \cite{ks}, the minimum value of $\vert 
f_{\rm NL}\vert$ that will become detectable from the analysis of MAP and 
{\it Planck} data, after properly subtracting detector noise and foreground 
contamination, is as large as $\sim ~20$, and $5$ respectively. 

In this paper we show that sizeable and detectable non-Gaussian
perturbations both in the adiabatic and the isocurvature modes 
naturally arise during inflation when extra scalar degrees of freedom are
present other than the inflaton field. In such a case, 
the adiabatic and isocurvature perturbations 
are  correlated \cite{L, goetal, bmr1, bmr2} as a result of 
an oscillation mechanism 
similar to the phenomenon leading to neutrino oscillations 
\cite{bmr1}$~$\footnote{This phenomenon was first described  in
Ref. \cite{grt} where it was pointed out that dangerous relics
may be  generated as coherent states through the oscillation
mechanism with the inflaton field in the preheating phase after inflation, 
leading to tight constraint on the reheating temperature.}.
This may happen, for instance, if the inflaton field is coupled 
to the 
other scalar degrees of freedom, as expected on particle physics grounds. 
If these scalar degrees of freedom have large self-interactions, their 
quantum fluctuations are intrinsically non-Gaussian.  
This non-Gaussianity is transferred to the inflaton sector 
through the oscillation mechanism
and is left imprinted in the adiabatic and isocurvature modes.   
We show that the CMB angular bispectrum is 
sourced not only by the intrinsic adiabatic and isocurvature bispectrum 
but also by cross-correlation terms, providing a characteristic 
and detectable signature of these non-Gaussian inflationary perturbations. 

The idea that an isocurvature perturbation mode produced during inflation 
could be non-Gaussian is certainly not new \cite{AGW, nongaussianinflation, 
LiM}. These scenarios, however, 
have the disadvantage that it is generally difficult to fit  
the observed pattern of CMB anisotropies in terms of isocurvature 
perturbations alone.  
The possibility of generically cross-correlating the adiabatic 
and isocurvature modes is attractive \cite{LR, BMT, TrottaetalAm} 
both because of its wider capability of 
reproducing the observed CMB angular power-spectrum and because of 
the possibility of introducing non-Gaussianity 
in the adiabatic mode too.\\
Moreover the characteristic signatures of these non-Gaussian inflationary perturbations could be also a way to break some degeneracies between the cosmological parameters which usually arise in scenarios where correlated adiabatic and isocurvature perturbations are present \cite{BMT, TrottaetalAm}. Indeed there exist other mechanisms to produce non-Gaussian primordial perturbations, such as single-field models allowing for an initial state which is not the ground state \cite{nonground} or, outside the inflationary paradigm, cosmic defects \cite{text}. The scenario we propose differs from the other mechanisms
since it can generate non-Gaussian perturbations which are adiabatic 
through the oscillation mechanism mentioned above.

The plan of the paper is as follows. In Section 2 we derive a general 
formula for the CMB angular bispectrum in models where correlated adiabatic and isocurvature perturbation modes are present. The physical mechanism 
by which these modes can be produced during inflation is summarized in 
Section 3. The resulting form for the angular bispectrum is obtained
in Section 4 in the case of adiabatic plus cold dark matter isocurvature perturbations. Section 5 contains our conclusions.

\section{The CMB Angular Bispectrum}

In order to  investigate possible non-Gaussian features of the 
CMB one can consider the  angular three point correlation function
\beq
\left \langle \frac{\Delta T}{T}({\bf{\hat{n}}}_1)  
\frac{\Delta T}{T}({\bf{\hat{n}}}_2)  
\frac{\Delta T}{T}({\bf{\hat{n}}}_3) \right \rangle  
=
\sum _{l_i,m_i} \langle a_{l_1m_1}a_{l_2m_2}a_{l_3m_3} \rangle Y_{l_1m_1}
({\bf{\hat{n}}}_1) Y_{l_2m_2} ({\bf{\hat{n}}}_2) Y_{l_3m_3} ({\bf{\hat{n}}}_3)
\, ,
\eeq
where hats denote unit 
vectors and  we have used the usual expansion 
of the CMB temperature anisotropy in spherical 
harmonics $Y_{lm} ({\bf{\hat{n}}})$ with coefficients
\beq
a_{lm}=\int d{\bf{\hat{n}}}\, Y^{*}_{lm}({\bf{\hat{n}}}) 
\frac{\Delta T}{T} ({\bf{\hat{n}}})\, .
\eeq
The angular CMB bispectrum is the 
harmonic conjugate of the three-point correlation function and is given by
\beq \label{bisp}
\langle a_{l_1m_1}a_{l_2m_2}a_{l_3m_3} \rangle = 
\left (\begin{array}{ccc}
l_1 & l_2 & l_3\\
m_1 & m_2 & m_3 
\end{array} \right) B_{l_1l_2l_3},
\eeq
where the first term is  the Wigner 3j symbol and $B_{l_1l_2l_3}$ is 
the angle-averaged bispectrum, which is the observational quantity. 

To  calculate the bispectrum one has to properly take into account
the  initial conditions  in the 
radiation dominated epoch after the end of inflation. Such initial
conditions  may  reflect either the adiabatic or
isocurvature nature of the
cosmological perturbations. In general, however, one expects
a mixture of adiabatic and isocurvature perturbations with
a nonvanishing cross-correlation \cite{L, goetal, bmr1, bmr2, LR, BMT, 
TrottaetalAm}. 

For \emph{pure} adiabatic perturbations the harmonic 
coefficients $a_{lm}$ are given by \cite{MaBert}
\beq \label{aa}
a_{lm} = 4\pi\, (-i)^l\, \int d^3k\, \hat{\Phi} (\textbf{k})\, \Delta_{l}(k) 
Y_{lm}^{*} (\bf{\hat{k}})\, ,
\eeq
where $\hat{\Phi} (\textbf{k})$  indicates 
the primordial gravitational potential 
perturbation and $\Delta_{l}(k)$ is the 
\emph{radiation}, or \emph{CMB, transfer function}.
In the large scale limit, one  recovers 
the Sachs-Wolfe 
effect 
\beq
\frac{\Delta T}{T} = \frac{1}{3}\Phi
\eeq
where $\Phi$ 
is the gravitational potential at
recombination, by choosing $\Delta_{l}(k) = 1/3\, j_{l}[k (\tau_{0} - 
\tau_{rec})]$, $\tau_0$ being the conformal time at present 
and $\tau_{rec}$ the conformal time at recombination.\\
In the case of  pure isocurvature perturbations one simply 
inserts the initial entropic perturbation $S(\textbf{k})$ in 
Eq.  (\ref{aa}) in place of the gravitational potential perturbation
(see, for example, \cite{bert}). Of course, this corresponds to  a different 
radiation transfer function which can be called $\Delta^{S}_{l}(k)$.\\
Having the expression for $a_{lm}$ it is possible to calculate the 
bispectrum. 
Following the formalism of Ref. \cite{wk}, one finds 
for pure adiabatic perturbations the following expression 
\begin{eqnarray} \label{ba}
\langle a_{l_1m_1}a_{l_2m_2}a_{l_3m_3} \rangle & = 
& (4\pi)^3 (-i)^{l_1+l_2+l_3} \int  
d^3k_1d^3k_2d^3k_3 \\ \nonumber
& & \times Y^*_{l_1m_1} ({\bf{\hat{k}}}_1) Y^*_{l_2m_2} ({\bf{\hat{k}}}_2) 
Y^*_{l_3m_3} ({\bf{\hat{k}}}_3) \\ \nonumber
& & \times \delta^3({\bf{k}}_1+{\bf{k}}_2+{\bf{k}}_3) 
P^{(3)}_{\hat{\Phi}}(k_1,k_2,k_3)\\ \nonumber
& & \times \Delta_{l_1}(k_1) \Delta_{l_2}(k_2) \Delta_{l_3}(k_3)
\end{eqnarray}
where
\beq 
\label{dd}
\langle \hat{\Phi}({\bf{k}}_1) \hat{\Phi}({\bf{k}}_2) \hat{\Phi}
({\bf{k}}_3) \rangle =
\delta^3({\bf{k}}_1+{\bf{k}}_2+{\bf{k}}_3)  P^{(3)}_{\hat{\Phi}}(k_1,k_2,k_3)
\eeq
is the three-dimensional bispectrum of the gravitational potential. A similar
expression holds for pure isocurvature perturbations. We now analyze 
what happens in the most general case in which both adiabatic and 
isocurvature modes are present and are correlated.

\subsection{Mixture of  adiabatic and entropy perturbations} 

In the case of initial adiabatic \emph{plus} entropy 
perturbations, we write the coefficient $a_{lm}$  as
\beq \label{ai}
a_{lm} = 4\pi\, (-i)^l\, \int d^3k\, \left[ \hat{\Phi} 
(\textbf{k})\, \Delta^{A}_{l}(k) + S(\textbf{k})\, \Delta^{S}_{l}(k) \right]\, 
 Y_{lm}^{*} (\bf{\hat{k}})\, ,
\eeq
where $\Delta^{A}_{l}(k)$ and $\Delta^{S}_{l}(k)$ are the tranfer functions 
for the adiabatic and the entropy perturbation modes, respectively. 
This expression is consistent with the fact that 
the equations for the evolution of cosmological perturbations are linear. 
As a check, one can consider the Sachs-Wolfe effect for adiabatic 
$(\hat{\Phi})$ plus cold dark matter isocurvature ($S_c$) 
perturbations \cite{LR}:
\begin{eqnarray} \label{DT}
\left( \frac{\Delta T}{T} \right)_{SW} & = & 
\left( \frac{\Delta T}{T} \right)_{AD} + 
\left( \frac{\Delta T}{T} \right)_{ISOC} \\ 
& = & \frac{1}{3}\, \Phi_{A} + 2\, \Phi_{S}\, . \nonumber
\end{eqnarray}
The first term on the r.h.s. of Eq. (\ref{DT}), containing the gravitational 
potential at large scales ($k \ll aH$) at the time of recombination, 
corresponds to  
the case of pure adiabatic perturbations. The second term corresponds to
the case of pure isocurvature perturbations. The two potentials are given by
\beq
\Phi_{A} = \frac{3}{10}\, \left( 3+\frac{4}{5} \Omega^{RD}_\nu \right) 
\hat{\Phi}\, , \qquad
\Phi_{S} = -\frac{1}{5}\, \Omega^{MD}_{c}\, S_c\, ,
\eeq
where  $\Omega^{MD}_{c}$ is the density parameter for 
the cold dark matter during the matter era
and $\Omega^{RD}_\nu$ the one for neutrinos 
during the radiation era. One can recover this result 
from Eq. (\ref{ai}) with the transfer functions 
$\Delta^{A}_{l}(k)= 1/3\, j_{l}(k \chi)$ and 
$\Delta^{S}_{l}(k)=-2/5\, j_l(k \chi) \Omega_{c}^{MD}$.\\
Note that the full transfer functions 
take into account all the other effects playing a role in 
the generation of the temperature anisotropies $\Delta T/T$ 
(such as the integrated Sachs-Wolfe effect 
emerging -- for example -- in the presence of a cosmological constant, and  
various small scale effects \cite{ks}).

Given the expression (\ref{ai}), 
if  adiabatic and  entropy perturbations are correlated 
we find for the bispectrum a result similar to Eq. (\ref{ba}), 
but with a more complicated structure
\begin{eqnarray} \label{b1}
\langle a_{l_1m_1}a_{l_2m_2}a_{l_3m_3} \rangle & = 
& (4\pi)^3 (-i)^{l_1+l_2+l_3} 
\int  d^3k_1d^3k_2d^3k_3 \\ \nonumber
& & \times Y^*_{l_1m_1} ({\bf{\hat{k}}}_1) Y^*_{l_2m_2} ({\bf{\hat{k}}}_2) 
Y^*_{l_3m_3} ({\bf{\hat{k}}}_3) 
\times \delta^3({\bf{k}}_1+{\bf{k}}_2+{\bf{k}}_3) \\ \nonumber
& & \times [P^{(3)}_{\hat{\Phi}}(k_1,k_2,k_3)\, 
\Delta^{A}_{l_1}(k_1) \Delta^{A}_{l_2}(k_2) \Delta^{A}_{l_3}(k_3) \\ \nonumber
& & +  P^{(3)}_S(k_1,k_2,k_3)\, \Delta^{S}_{l_1}(k_1) 
\Delta^{S}_{l_2}(k_2) \Delta^{S}_{l_3}(k_3) \\ \nonumber
& & +  P^{(3)}_{AAS}(k_1,k_2,k_3)\, \Delta^{A}_{l_1}(k_1) 
\Delta^{A}_{l_2}(k_2) \Delta^{S}_{l_3}(k_3) \\ \nonumber
& & + (A, S, A) + (S, A, A) + (S, S, A) + (S, A, S) + (A, S, S)]\, .
\end{eqnarray}
As expected, the  bispectrum gets contributions from  the
adiabatic modes (\ref{dd}), from  the isocurvature modes
\beq 
\langle S({\bf{k}}_1) S({\bf{k}}_2) S({\bf{k}}_3) \rangle =
\delta^3({\bf{k}}_1+{\bf{k}}_2+{\bf{k}}_3)  P^{(3)}_S(k_1,k_2,k_3)
\eeq
and from the terms parametrizing the cross-correlation
between adiabatic and isocurvature modes, for example
\beq \label{mix}
\langle \hat{\Phi}({\bf{k}}_1) S({\bf{k}}_2) \hat{\Phi}({\bf{k}}_3) 
\rangle =
\delta^3({\bf{k}}_1+{\bf{k}}_2+{\bf{k}}_3)  P^{(3)}_{ASA}(k_1,k_2,k_3)
\eeq
where we have adopted the notation 
\beq
(A, S, A) \equiv P^{(3)}_{ASA}(k_1,k_2,k_3)\, 
\Delta^{A}_{l_1}(k_1) \Delta^{S}_{l_2}(k_2) 
\Delta^{A}_{l_3}(k_3)\, .
\eeq
Performing the angular integration following 
Ref. \cite{wk},  we obtain Eq. (\ref{bisp}), where 
\begin{eqnarray} \label{aab}
B_{l_1l_2l_3} & = & (8\pi)^3 \sqrt{\frac{(2l_1+1)(2l_2+1)(2l_3+1)}{4\pi}} 
\left (\begin{array}{ccc}
l_1 & l_2 & l_3 \\
0 & 0 & 0  \end{array} \right)\\
& \times & \int dk_1\, k_1^2\, dk_2\, k_2^2\, dk_3\, k_3^2\,  
J_{l_1l_2l_3}(k_1,k_2,k_3) \times  \nonumber \\
&  & [P^{(3)}_{\hat{\Phi}}(k_1,k_2,k_3)\, 
\Delta^{A}_{l_1}(k_1) \Delta^{A}_{l_2}(k_2) \Delta^{A}_{l_3}(k_3)\, + 
 \nonumber \\
 &  &  P^{(3)}_S(k_1,k_2,k_3)\, \Delta^{S}_{l_1}(k_1) 
\Delta^{S}_{l_2}(k_2) \Delta^{S}_{l_3}(k_3)\, +  \nonumber \\
 &  &  P^{(3)}_{AAS}(k_1,k_2,k_3)\, \Delta^{A}_{l_1}(k_1) 
\Delta^{A}_{l_2}(k_2) \Delta^{S}_{l_3}(k_3)\, +  \nonumber \\
 &  & (A, S, A) + (S, A, A) + (S, S, A) + (S, A, S) + (A, S, S)] \nonumber\, .
\end{eqnarray}
Note that the integral in Eq. (\ref{aab}) is proportional to the 
\emph{reduced} bispectrum defined in Ref. \cite{ks}. 
Indeed it  contains all the physical information on the bispectrum.

Our goal is now to show that large contributions
to the bispectrum (\ref{b1}) may naturally arise
when adiabatic and isocurvature modes are correlated.

\section{Adiabatic and entropy perturbations from inflation}

Correlated adiabatic and isocurvature modes  can be generated  during a 
period of inflation in which several scalar fields are present
\cite{L, goetal, bmr1, bmr2}. 
Indeed, adiabatic (curvature) perturbations are produced 
during a period of cosmological inflation 
that is driven by a single scalar field, the inflaton. 
On particle
physics grounds -- though -- it is natural to expect that 
this scalar field is coupled to other scalar degrees of freedom. 
This gives rise to oscillations between
the perturbation of the inflaton field and the perturbations of the 
other scalar degrees of freedom, similar to the 
phenomenon of neutrino oscillations. The  crucial 
observation  is that -- since the
degree of mixing is governed by the squared mass 
matrix of the scalar fields -- the oscillations can occur 
even if the energy density of the extra scalar fields
is much smaller than the energy density of the inflaton. 
The probability of oscillation is resonantly amplified 
when perturbations cross the horizon and the
perturbations in the inflaton field may disappear at 
horizon crossing giving rise to 
perturbations in scalar fields other than the inflaton. Adiabatic and
isocurvature perturbations are inevitably correlated 
at the end of inflation \cite{bmr1,bmr2}. 

It is exactly this strong correlation which
may give rise to large non-Gaussian features in the CMB anisotropy spectrum.
This is a simple, but important point. Gaussian features in the
CMB anisotropies are usually
expected in inflationary models because the inflaton potential
is required to be very flat. This amounts to saying that the 
interaction terms in the inflaton potential are present, but
small and non-Gaussian features are suppressed since the non-linearities 
in the inflaton potential are suppressed too. On the other hand, nothing 
prevents the inflaton field from being coupled to another scalar degree 
of freedom
whose energy density is much smaller than the one stored in the inflaton
field. It is natural
to expect that the  the self-interactions of such extra field 
or the interaction terms with the inflaton field are sizeable and
 they represent 
potential non-linear sources for non-Gaussianity. 
If during the inflationary epoch,  oscillations between
the perturbation of the inflaton field and the perturbations of the 
other scalar degrees of freedom occur, the non-Gaussian features generated
in the system of the extra field are efficiently communicated to the inflaton
sector and may be left imprinted in the CMB anisotropies.

Let us consider for simplicity the case of two scalar fields $\phi$ and $\chi$ 
interacting through a generic potential $V(\phi, \chi)$. 
The study of the field fluctuations $\delta \phi$ and $\delta \chi$ 
can be done using the Sasaki-Mukhanov variables\footnote{To simplify 
the calculation of the three-point correlation functions  one 
can reduce to a particular gauge, such as the 
spatially flat gauge $(\psi=0)$ in which the $Q_I$ variables 
concide with $\delta \phi_I$.} \cite{sm}
\beq \label{SM}
 Q_I\equiv \delta \phi_I +\frac{\dot{\phi_I}}{H}\, \psi       
\eeq
where $I=1,2$ with $\delta \phi_1=\delta \phi$, $\delta \phi_2=\delta \chi$ 
and  $\psi$ is the linear perturbation in the line element of the metric
\beq
ds^2 = -(1+2A)dt^2+2aB_idx^i dt+a^2[(1-2\psi)\delta_{ij}+2E_{ij}]dx^i dx^j.
\eeq
Using such  variables it is possible to define the adiabatic 
and entropy fields 
$Q_A$ and $\delta s$  
in terms of the original field perturbations $Q_{\phi}$ and $Q_{\chi}$ 
\cite{goetal}
\beq \label{dA}
Q_A=(\cos \beta)Q_\phi+(\sin \beta)Q_\chi \label{A}\, ,
\eeq
\beq \label{ds}
\delta s=(\cos \beta)Q_\chi-(\sin \beta)Q_\phi \label{s}\, ,
\eeq
where
\beq \label{angledef}
\cos \beta \equiv c_{\beta}=\frac{\dot{\phi}}
{\sqrt{\dot{\phi}^{2}+\dot{\chi}^{2}}}\, , 
\qquad  
\sin \beta \equiv s_{\beta}=\frac{\dot{\chi}}
{\sqrt{\dot{\phi}^{2}+\dot{\chi}^{2}}}\, ,
\eeq
and the dots stand for the derivatives with respect to the
cosmic time $t$.\\
The curvature perturbation \cite{CURV}
\beq
\label{def:calR} 
{\cal R}= H  \sum_I \left(\frac{\dot\varphi_I}{\sum_{J=1}^{N} 
\dot\varphi_J^2} \right) Q_{I} 
\eeq
deep in the radiation era can be written in terms of 
the adiabatic field $Q_A$\beq \label{R}
{\mathcal{R}}_{rad}= \frac{H}{c_\beta \dot{\phi} + s_\beta \dot{\chi}} 
Q_A \, 
\eeq 
where the r.h.s of this equation is evaluated at the end of inflation.\\
Let us now introduce the  slow-roll parameters for the 
two scalar fields $\phi$ and $\chi$
\beq
\epsilon_{I} =  \frac{M^{2}_{Pl}}{16\pi} \left( \frac{V_{\phi_I}}{V} 
\right)^2 \quad \textnormal{and} \quad \eta_{IJ}=\frac{M^{2}_{Pl}}{8\pi} 
\frac{V_{\phi_{I}\phi_{J}}}{V}\, ,  
\eeq
where $M_{Pl}$ is the Planck mass, $V_{\phi_I}=\partial V/ \partial 
\phi_I$, and $\phi_I=\phi,\chi$.\\
Having a sucessfull period of inflation requires 
that the potential is flat enough, that is $\epsilon_{I}$ and 
$\arrowvert\eta_{IJ} \arrowvert \ll 1$.
Now, making an expansion in the slow roll parameters to lowest order, 
it is possibile to write  the gravitational potential 
$\hat{\Phi}$ as \cite{bmr1, bmr2} 
\beq \label{1}
\hat{\Phi} = \frac{2}{3} {\cal{R}}_{rad} = 
\frac{2}{3} \frac{\sqrt{4 \pi}}{M_{Pl}} \frac{1}{\sqrt{\epsilon_{tot}}}\, 
Q_A\, 
\eeq
where  $\epsilon_{tot} = \epsilon_\phi + \epsilon_\chi$.

Under the hypothesis that the scalar field $\phi$ decays into 
``ordinary'' matter (the present day photons, neutrinos and baryons), 
while the scalar field $\chi$ decays only into cold dark matter 
(or it does not decay at all, like in the case of the superheavy dark matter
\cite{shdm}), an adiabatic 
($\hat{\Phi}$) and a cold dark matter isocurvature mode ($S_c$) 
will be generated in the post-inflationary epoch. 
To lowest order in the slow roll parameters, they are given 
by  expression (\ref{1}) and \cite{bmr1, bmr2}
\beq \label{2}
S_{c} = -3\ \frac{\sqrt{4\pi}}{m_{Pl}}\ \frac{\sqrt{\epsilon_{tot}}}
{(\pm \sqrt{\epsilon_\phi})(\pm \sqrt{\epsilon_\chi})} \delta s
\eeq
where the r.h.s.  
is  evaluated at the end of inflation as a matching condition. 
In a short-hand notation we  can write
\beq \label{conc}
\hat{\Phi} = A_0\, Q_A, \qquad  S = S_0\, \delta s,
\eeq
where $A_0$ and $S_0$ are just the ``amplitudes'' of $\hat{\Phi}$ and $S_c$.

As we will show in the next section, since it is quite 
natural to expect a nonzero correlation between 
the adiabatic field $Q_A$ and the entropy field $\delta s$  
generated during inflation \cite{bmr1}, non-Gaussian features
in the CMB anisotropies may be left imprinted.

\section{Primordial non-Gaussianity from inflation}

We are now in the position of 
relating the bispectrum in Eq. (\ref{b1}) with the expressions 
for $\hat{\Phi}$ and $S_c$ originated during a  a period of inflation.

Consider, for example, the $\langle \hat{\Phi}({\bf{k}}_1) 
\hat{\Phi}({\bf{k}}_2) \hat{\Phi}({\bf{k}}_3) 
\rangle$ term. One finds
\begin{eqnarray}
\label{ddd}
\langle \hat{\Phi}({\bf{k}}_1) \hat{\Phi}({\bf{k}}_2) \hat{\Phi}({\bf{k}}_3) 
\rangle & = & A_{0}^3 \langle Q_A({\bf{k}}_1)  
Q_A({\bf{k}}_2)  Q_A({\bf{k}}_3) \rangle \nonumber\\ 
& = & A_{0}^3 \langle (c_\beta Q_{\phi1} + s_\beta Q_{\chi1}) 
(c_\beta Q_{\phi2} + s_\beta Q_{\chi2}) (c_\beta Q_{\phi3} 
+ s_\beta Q_{\chi3}) \rangle \\ \nonumber
& = & A_{0}^3\,  [c_\beta^3 \, \langle Q_{\phi1} Q_{\phi2} Q_{\phi3} 
\rangle + c_\beta^2 s_\beta \langle Q_{\phi1} Q_{\phi2} Q_{\chi3} 
\rangle + c_\beta^2 s_\beta \langle Q_{\phi1} Q_{\chi2} Q_{\phi3} \rangle 
\nonumber\\
& & + c_\beta s_\beta^2 \langle Q_{\phi1} Q_{\chi2} Q_{\chi3} \rangle + 
s_\beta c_\beta^2 \langle Q_{\chi1} Q_{\phi2} Q_{\phi3} \rangle + 
s_\beta^2 c_\beta \langle Q_{\chi1} Q_{\phi2} Q_{\chi3} \rangle  \nonumber \\
& & + s_\beta^2 c_\beta \langle Q_{\chi1} Q_{\chi2} Q_{\phi3} 
\rangle + s_\beta^3 \langle Q_{\chi1} Q_{\chi2} Q_{\chi3} \rangle ]\, \nonumber      
\end{eqnarray}
where, for example,  $Q_{\phi1}$
 stands for $Q_{\phi}({\bf{k}}_1)$ and we have used Eq.
 (\ref{dA}). Analogous expressions hold for the remaining terms. 

Our goal is now  to show that a large amount of non-Gaussianity can be 
generated in the presence of correlated adiabatic and entropy perturbations. 
First of all, we note that the bispectrum is a sum of different  
three-point correlation functions. The coefficients in front of each
correlation function involve  mixing angles which parametrize the amount of 
mixing between
the adiabatic and the isocurvature modes. If such mixing is sizeable,
all coefficients are of order unity and one expects that nonlinearities
in the perturbation of the scalar field $\chi$ may be efficiently
transferred to the inflaton sector, thus generating large
non-Gaussian features.

Because expressions are quite lengthy and might
obscure our  point, from now on and just for
illustrative purposes we make some simplifying 
hypothesis and 
assume that the dominant nonlinear terms are those sourced by   
the three-point correlation function  
$\langle Q_{\chi1} Q_{\chi2} Q_{\chi3} \rangle$. This could be the 
case for a Langrangian of scalar fields $\phi$ and $\chi$ in which 
the largest  coupling is for the   
$\frac{\mu}{3} \chi^3$-term. Let us also assume that
the field $\chi$ is lighter than the Hubble rate during inflation. 

Under these assumptions, we can rewrite the bispectrum  ({\ref{b1}}) as
\begin{eqnarray} 
\label{new}
\langle a_{l_1m_1}a_{l_2m_2}a_{l_3m_3} \rangle & = & (4\pi)^3 
(-i)^{l_1+l_2+l_3} 
\int  d^3k_1d^3k_2d^3k_3 \\ \nonumber
& & \times Y^*_{l_1m_1} ({\bf{\hat{k}}}_1) Y^*_{l_2m_2} ({\bf{\hat{k}}}_2) 
Y^*_{l_3m_3} ({\bf{\hat{k}}}_3) 
\times \delta^3({\bf{k}}_1+{\bf{k}}_2+{\bf{k}}_3) \\ \nonumber
& & \times \{ A_0^3 s_\beta^3\,  P^{(3)}_{Q_\chi}(k_1,k_2,k_3) 
\Delta^{A}_{l_1}(k_1) \Delta^{A}_{l_2}(k_2) \Delta^{A}_{l_3}(k_3)  
\\ \nonumber
& & + S_{0}^3 c_\beta^3\,  P^{(3)}_{Q_\chi}(k_1,k_2,k_3)  
\Delta^{S}_{l_1}(k_1) \Delta^{S}_{l_2}(k_2) 
\Delta^{S}_{l_3}(k_3)  \\ \nonumber
& & +
A_0^2 S_0 s_\beta^2 c_\beta\,  P^{(3)}_{Q_\chi}
(k_1,k_2,k_3)[ \Delta^{A}_{l_1}(k_1) \Delta^{A}_{l_2}(k_2) 
\Delta^{S}_{l_3}(k_3) \\ \nonumber 
& & + (k_1 \leftrightarrow k_3; l_1 \leftrightarrow l_3) + 
(k_2 \leftrightarrow k_3; l_2 \leftrightarrow l_3)] \\ \nonumber
& & +S_0^2 A_0 c_\beta^2 s_\beta\,  P^{(3)}_{Q_\chi}(k_1,k_2,k_3) 
[\Delta^{S}_{l_1}(k_1) \Delta^{S}_{l_2}(k_2) \Delta^{A}_{l_3}(k_3)
 \\ \nonumber 
& & + (k_1 \leftrightarrow k_3; l_1 \leftrightarrow l_3) + (k_2 
\leftrightarrow k_3; l_2 \leftrightarrow l_3)]\,    \}\, 
\end{eqnarray}
where 
\beq 
\langle Q_{\chi1} Q_{\chi2} Q_{\chi3} \rangle =  
\delta^3({\bf{k}}_1+{\bf{k}}_2+{\bf{k}}_3)  P^{(3)}_{Q_\chi}(k_1,k_2,k_3)\, .
\eeq
The angular part of the integral can be calculated as done in subsection 2.1.

The next step is to further  reduce the  expression for the 
bispectrum by expanding  $\langle Q_{\chi1} Q_{\chi2} Q_{\chi3} \rangle$.
This is necessary because, in the
presence of large mixing, $Q_\phi$ and $Q_\chi$ are
not ''mass-eigenstates'' of the system, but just interaction eigenstates.
The situation here is analogous to what happens for light neutrinos
where the three different flavors of neutrinos represent interaction
eigenstates, but they do not represent 
mass-eigenstates because of the  mixing among the flavors giving rise
to the phenomenon of neutrino oscillations.

We first define the comoving fields  $\tilde{Q}_\phi = a Q_\phi$ and 
$\tilde{Q}_\chi= a Q_\chi$ and   then we introduce a 
basis for annihilation and 
creation operators $a_i$ and $a_i^{\dag}$  \cite{bmr1}. 
We perform the decomposition 
($\tau$ is the conformal time):
\begin{eqnarray}
\label{deco}
\left(
\begin{array}{c}
\widetilde{Q}_\phi\\
\widetilde{Q}_\chi\end{array}
\right)
&=&\U\,\int\,\frac{d^3k}{(2\pi)^{3/2}}\,\left[e^{i{\bf k}\cdot{\bf x}}\,
h(\tau)\left(
\begin{array}{c}
a_1(k)\\
a_2(k)\end{array}\right)\,+\,{\rm h.c.}\right]\, ,\nonumber\\
\left(
\begin{array}{c}
\Pi_{\widetilde{Q}_\phi}\\
\Pi_{\widetilde{Q}_\chi}\end{array}
\right)
&=&\U\,\int\,\frac{d^3k}{(2\pi)^{3/2}}\,\left[e^{i{\bf k}\cdot{\bf x}}\,
\widetilde{h}(\tau)\left(
\begin{array}{c}
a_1(k)\\
a_2(k)\end{array}\right)\,+\,{\rm h.c.}\right]\, ,
\end{eqnarray}
where $\Pi_{\widetilde{Q}_\phi}$ and 
$\Pi_{\widetilde{Q}_\chi}$ are the conjugate momenta
of $\widetilde{Q}_\phi$ and 
$\widetilde{Q}_\chi$ respectively, and $h$ and $\widetilde{h}$ are 
two $2\times 2$ matrices  satisfying the relation
\beq
\left[h\,\widetilde{h}^* -h^*\,\widetilde{h}^T\right]_{ij}=
i\,\delta_{ij},
\eeq
derived from the canonical quantization condition. 

The matrix $\U$ is a rotation matrix    
\beq
\label{u}
\U=\left(\begin{array}{cc}
\cos\theta & -\sin\theta\\
\sin\theta & \cos\theta
\end{array}\right)
\eeq
which diagonalizes the squared mass matrix of the two scalar 
field perturbations $Q_{\phi}$ and $Q_{\chi}$ 
\beq
\label{mm}
{\cal M}^2_{IJ}= V_{\phi_I\phi_J} - 8\pi/M_{Pl}^2 a^3\, 
\left( a^3/H\,\, \dot \phi_I \dot \phi_J \right)^{\cdot}\simeq
\frac{8\pi V}{M_{Pl}^2}\left[ \eta_{IJ}-2\, (\pm \sqrt{\epsilon_{I}})
(\pm \sqrt{\epsilon_{J}})\right]
\, ,
\eeq
where the sign $\pm$ stands for the cases $\dot{\phi_I}(\dot{\phi_J}) > 0$ 
and $< 0$ respectively. 
The mixing angle $\theta$ is given by
\beq
\tan 2\theta=\frac{2\,{\cal M}^2_{\chi\phi}}{{\cal M}^2_{\phi\phi}-
{\cal M}^2_{\chi\chi}}\, .
\eeq
One can  envisage different situations:

{\it i)} Inflation is driven by the inflaton field $\phi$ and
there is another scalar field $\chi$ with a simple polynomial 
potential $V(\chi)\propto \chi^n$ leading to a vacuum expectation value
$\langle\chi\rangle =0$. In such a case, $\sin\beta=\sin\theta=0$ and there 
is no mixing between the inflaton field and the $\chi$-field
as well as no cross-correlation between the adiabatic and isocurvature
modes. Nonvanishing non-Gaussianity will be present in the isocurvature
mode. This is indeed a known result \cite{AGW, LiM}.
Non-Gaussian adiabatic perturbations may also arise if
the $\chi$-field decays late after inflation \cite{LiM, LW}.     

{\it ii)} Inflation is driven by two scalar fields $\phi$ and $\chi$
with equal mass, $V=\frac{m^2}{2}(\phi^2+\chi^2)$. In such a case the
mixing is maximal, $\beta=
\theta=\pi/4$. Nevertheless, the cross-correlation is again vanishing
\cite{goetal,bmr1,bmr2} and the bispecrum gets contributions from 
adiabatic and isocurvature modes independently, since in this case the 
terms parametrizing the cross-correlation in Eq. (\ref{b1}) vanish.  A  
term $\frac{\mu}{3}\chi^3$ in the Lagrangian would be a source of 
non-Gaussianity and at the same time it would switch on a cross 
correlation between the adiabatic and the isocurvature modes, thus
 producing nonzero cross terms in Eq. (\ref{new}). However, 
these non-Gaussianities would be small because of slow-roll
conditions. 

{\it iii)} Inflation is driven by an inflaton field $\phi$ 
and there is another scalar field $\chi$ whose vacuum expectation value 
depends on the inflaton field and -- eventually -- on the Hubble parameter
$H$ and some other mass scale $\mu$, $\langle \chi\rangle=f(\phi,H,\mu)$.
Under these circumstances, $\langle
\dot\chi\rangle=\partial f/\partial \phi\,  \dot\phi+
\partial f/\partial H\,  \dot H$. As in illustrative case, let us 
restrict ourselves 
to the case in which  $\partial f/\partial \phi\,  \dot\phi$ is 
the dominant term and  we can approximate
$\langle\dot\chi\rangle=\partial f/\partial \phi \dot\phi$. 
We have therefore
$\tan\beta\simeq \partial f/\partial \phi $ and $\dot\beta\simeq 
(\partial f/\partial \phi)^{\cdot} / [1+ (\partial f/\partial \phi)^2]$.
In such a case, cross-correlation  between the adiabatic and the 
isocurvature modes may be large and non-Gaussianity
may be efficiently transferred from one mode to the other. 

We can now reduce  $\langle Q_{\chi1} Q_{\chi2} Q_{\chi3} \rangle$ 
using the decomposition (\ref{deco}) and making some further 
approximations justified if slow-roll conditions are attained.
In fact, using a perturbative method, it can be checked that the 
contributions to  $\langle Q_{\chi1} Q_{\chi2} Q_{\chi3} \rangle$ coming 
from terms proportional to the non-diagonal elements $h_{12}$ and $h_{21}$ 
can be neglected since  $h_{12}$ and $h_{21}$ are ${\mathcal{O}}(\epsilon_I, 
\eta_{IJ})$, 
and on superhorizon scales $k \ll aH$  $h_{11}$
and $h_{22}$ are  Hankel functions \cite{bmr1}. Thus we can 
neglect the non diagonal terms of the $h$ matrix and 
 write\footnote{In such a case the system is diagonalized by the matrix 
$\U$.}
\begin{eqnarray} \label{rotation}
\tilde{Q}_\chi({\bf{k}}) & = & [s_\theta h_{11}\, a_1({\bf{k}}) + 
c_\theta h_{22}\, a_2({\bf{k}})] + [s_\theta h_{11}^*\,
 a_1^{\dag}(-{\bf{k}}) + c_\theta h_{22}^*\, a_2^{\dag}(-{\bf{k}})] \\ 
\nonumber
& = & s_\theta [h_{11}\, a_1({\bf{k}}) + h_{11}^*\, a_1^{\dag}(-{\bf{k}})] 
+ c_\theta  [h_{22}\, a_2({\bf{k}}) + h_{22}^*\, a_2^{\dag}(-{\bf{k}})] \\ 
\nonumber
& \equiv & s_\theta h_1\, I_1 + c_\theta h_2\, I_2\, , 
\end{eqnarray}
where, for simplicity of notation, we have indicated the diagonal terms of the $h$ matrix as $h_{11} \equiv h_1\, , h_{22} \equiv h_2$ and we have defined  two new fields $I_1$ and $I_2$ 
just by collecting the functions $h_{11}$ and $h_{22}$. 
After all these manipulations we arrive at three-point function  
\begin{eqnarray} \label{I} \nonumber
\langle \tilde{Q}_{\chi1} \tilde{Q}_{\chi2} \tilde{Q}_{\chi3} \rangle & = &
\langle (s_\theta\, h_{1\bf{1}}\, I_{1\bf{1}} + c_\theta\, h_{2\bf{1}}\, I_{2\bf{1}})  
(s_\theta\, h_{1\bf{2}}\, I_{1\bf{2}} + c_\theta\, h_{2\bf{2}}\, I_{2\bf{2}})  
(s_\theta\, h_{1\bf{3}}\, I_{1\bf{3}} + c_\theta\, h_{2\bf{3}}\, I_{2\bf{3}}) \rangle \\ 
& = & s_\theta^3\, (h_{1\bf{1}}\, h_{1\bf{2}}\, h_{1\bf{3}}) \langle  I_{1\bf{1}}\, I_{1\bf{2}}\, I_{1\bf{3}} 
\rangle + s_\theta^2 c_\theta\, (h_{1\bf{1}}\, h_{1\bf{2}}\, h_{2\bf{3}})  \langle  I_{1\bf{1}}\, 
I_{1\bf{2}}\, I_{2\bf{3}} \rangle \\ \nonumber
& & + s_\theta^2 c_\theta\,  (h_{1\bf{1}}\, h_{2\bf{2}}\, h_{1\bf{3}})  \langle  I_{1\bf{1}}\, I_{2\bf{2}}\,
 I_{1\bf{3}} \rangle + s_\theta 
c_\theta^2\,  (h_{1\bf{1}}\, h_{2\bf{2}}\, h_{2\bf{3}}) \langle  I_{1\bf{1}}\, I_{2\bf{2}}\, I_{2\bf{3}} \rangle \\ \nonumber
& & + c_\theta s_\theta^2\,  (h_{2\bf{1}}\, h_{1\bf{2}}\, h_{1\bf{3}}) \langle  
I_{2\bf{1}}, I_{1\bf{2}}\, I_{1\bf{3}} \rangle + c_\theta^2 s_\theta\,  (h_{2\bf{1}}\, 
h_{1\bf{2}} h_{2\bf{3}}) \langle  I_{2\bf{1}}\, I_{1\bf{2}}\, I_{2\bf{3}} \rangle \\ \nonumber
& & + c_\theta^2 s_\theta\,  (h_{2\bf{1}}\, h_{2\bf{2}}\, h_{1\bf{3}}) \langle  I_{2\bf{1}}\, I_{2\bf{2}}\, 
I_{1\bf{3}} \rangle + c_\theta^3\,  (h_{2\bf{1}}\, h_{2\bf{2}}\, h_{2\bf{3}}) \langle  I_{2\bf{1}}\, I_{2\bf{2}}\, 
I_{2\bf{3}} \rangle\, ,        
\end{eqnarray}
where the indices `` ${\bf{1,2,3}}$ '', as usual, indicate that the quantities are evaluated at
${\bf{k}}_1, {\bf{k}}_2$ and ${\bf{k}}_3$.\\
If we indicate 
\beq
\langle I_{i\bf{1}}\, I_{j\bf{2}}\, I_{k\bf{3}} \rangle =  
\delta^3({\bf{k}}_1+{\bf{k}}_2+{\bf{k}}_3)\, 
P^3_{ijk}(k_1,k_2,k_3), \, \, i,j,k = 1,2
\eeq
because of rotation and translation invariance, 
it is easy to convince oneself that 
the terms  $\langle I_{i\bf{1}}\, I_{j\bf{2}}\, I_{k\bf{3}} \rangle$ are 
invariant under an exchange of $ {\bf{k}}_1,{\bf{k}}_2,{\bf{k}}_3$. 
Thus, taking into account that the operators $I_1$ and $I_2$ commute, 
one can check that, for example,
$ \langle  I_{1\bf{1}}\, I_{1\bf{2}}\, I_{2\bf{3}} \rangle$ is equal to 
$ \langle  I_{1\bf{1}}\, I_{2\bf{2}}\, I_{1\bf{3}} \rangle$.
The expression (\ref{I}) is thus  further simplified to
\begin{eqnarray} \label{II} 
\langle \tilde{Q}_{\chi1} \tilde{Q}_{\chi2} \tilde{Q}_{\chi3} \rangle & = &
s_\theta^3\, (h_{1\bf{1}}\, h_{1\bf{2}}\, h_{1\bf{3}}) \langle  I_{1\bf{1}}\, I_{1\bf{2}}\, 
I_{1\bf{3}} \rangle
 + c_\theta^3\,  (h_{2\bf{1}}\, h_{2\bf{2}}\, h_{2\bf{3}}) \langle  I_{2\bf{1}}\, I_{2\bf{2}}
\, I_{2\bf{3}}
 \rangle \nonumber \\
& & + s_\theta^2 c_\theta\,  \langle  I_{1\bf{1}}\, I_{1\bf{2}}\, I_{2\bf{3}} \rangle  
(h_{1\bf{1}}\, h_{1\bf{2}}\, h_{2\bf{3}} + h_{1\bf{1}}\, h_{2\bf{2}}\, h_{1\bf{3}}
+ h_{2\bf{1}}\, h_{1\bf{2}}\, h_{1\bf{3}}) 
\nonumber \\
& & + c_\theta^2 s_\theta\,  \langle  I_{1\bf{1}}\, I_{2\bf{2}}\, I_{2\bf{3}} \rangle \  
(h_{1\bf{1}}\, h_{2\bf{2}}, h_{2\bf{3}} + h_{2\bf{1}}\, h_{1\bf{2}}\, h_{2\bf{3}} 
+ h_{2\bf{1}}\, h_{2\bf{2}}\, h_{1\bf{3}})\, .   
\end{eqnarray}

In order to make a quantitative estimate of the three-point function  
(\ref{II}), we first notice that the coefficients in front of the various 
terms are of order unity, provided the degree of mixing is large.  
We can borrow the expression for each three-point function appearing in 
Eq. (\ref{II}) from the calculation of \cite{FRS}, which is done for 
an effectively massless scalar field $\chi$ 
with  cubic self-interactions. In the de Sitter background it is given 
by 
\begin{equation}
\langle(\chi({\bf k}_1)\chi({\bf k}_2)\chi({\bf k}_3)\rangle = 
\frac{1}{6} \mu H^2 (k_1 k_2 k_3)^{-3} F(k_1,k_2,k_3) 
\delta^{(3)}({\bf k}_1 + {\bf k}_2 + {\bf k}_3) \;,  
\end{equation}  
where 
\begin{equation}
F(k_1,k_2,k_3) \simeq - \beta (k_1^3 + k_2^3 + k_3^3) 
\end{equation}
and, for instance, $\beta \sim 60$ if one is interested in the scales 
relevant for large-angle CMB anisotropies. 
Notice that the functional form of the bispectrum is the same 
found by Gangui et al. \cite{Getal}, who made different assumptions 
and used the stochastic approach to inflation. 

Plugging the above expressions into the CMB angular bispectrum one gets 
the standard relation \cite{Getal, Vetal} giving  
the angular bispectrum $B_{l_1 l_2 l_3}$ as a sum of products of two angular 
power-spectra, $C_{l_i} C_{l_j}$. The non-Gaussianity 
amplitude is monitored by the dimensionless strength 
$f_{\rm NL} = {\cal O}(\mu/H)$. 

\section{Conclusions}

In this paper we have studied inflationary models where extra scalar degrees 
of freedom other than the inflaton exist. This allows isocurvature 
perturbation modes to be switched on during the inflationary evolution 
besides the usual adiabatic one. As previously shown  
\cite{goetal, bmr1, bmr2}, a generic prediction of these models is that 
non-zero cross-correlations arise among adiabatic and isocurvature 
fluctuations. 
Here we exploited this physical process as an efficient tool to transfer 
non-Gaussian features from the isocurvature to the adiabatic mode. 
Sizeable non-Gaussianity can be easily produced in the 
non-inflatonic sector, by self-interactions leading to non-linearities 
in their evolution. This is because, unlike the inflaton case, the 
self-interaction strength in such an extra scalar sector does not suffer 
from the usual slow-roll conditions. 
In order to make use of our results for practical purposes, 
one might introduce a simple non-Gaussian model. For instance, one can 
parametrize the non-Gaussianity in 
the peculiar gravitational potential as 
\begin{equation} 
\Phi = \varphi_1 + f_{\rm NL} (\varphi_2^2 - \langle \varphi_2^2 \rangle) 
+ {\cal O} (f_{\rm NL}^2) 
\end{equation}
(and a similar expression for the entropy mode), where 
$\varphi_1$ and $\varphi_2$ are zero-mean Gaussian fields with non-zero 
cross-correlation $\langle \varphi_1 \varphi_2 \rangle \neq 0$. 
The non-Gaussianity strength $f_{\rm NL}$, being sourced by  
the non-inflatonic scalar sector of the theory, is not generally 
constrained by the slow-roll conditions of inflationary dynamics.  
This may make the non-Gaussian signatures accessible by future CMB 
satellite experiments. 

\vskip1cm
\centerline{\large\bf Acknowledgements} 
\vskip 0.2cm
NB acknowledges the Marie Curie Fellowship of the European Community Program 
{\it Human Potential} under contract N. HPNT-CT-2000-00096.

\end{document}